\documentclass[aps,prl,twocolumn,showpacs,superscriptaddress]{revtex4}

\usepackage {graphicx}
\usepackage{epsf}

\begin{document}

\title{Atomic site sensitive processes in low energy ion-dimer collisions}

\author{W. Iskandar}
\affiliation{CIMAP, CEA - CNRS - ENSICAEN, BP 5133, F-14070, Caen cedex 5, France}
\author{J. Matsumoto}
\affiliation{Department of Chemistry, Tokyo Metropolitan University, 1-1 Minamiosawa, Hachiouji-shi, Tokyo 192-0397, Japan}
\author{A. Leredde}
\affiliation{Physics Division, Argonne National Laboratory, Argonne, Illinois 60439, USALPC}
\author{X. Fl\'echard}
\email{flechard@lpccaen.in2p3.fr}
\affiliation{LPC Caen, ENSICAEN, Universit\'e de Caen, CNRS/IN2P3, Caen, France}
\author{B. Gervais}
\affiliation{CIMAP, CEA - CNRS - ENSICAEN, BP 5133, F-14070, Caen cedex 5, France}
\author{S. Guillous}
\affiliation{CIMAP, CEA - CNRS - ENSICAEN, BP 5133, F-14070, Caen cedex 5, France}
\author{D. Hennecart}
\affiliation{CIMAP, CEA - CNRS - ENSICAEN, BP 5133, F-14070, Caen cedex 5, France}
\author{A.M\'ery}
\affiliation{CIMAP, CEA - CNRS - ENSICAEN, BP 5133, F-14070, Caen cedex 5, France}
\author{J. Rangama}
\affiliation{CIMAP, CEA - CNRS - ENSICAEN, BP 5133, F-14070, Caen cedex 5, France}
\author{C.L. Zhou}
\affiliation{CIMAP, CEA - CNRS - ENSICAEN, BP 5133, F-14070, Caen cedex 5, France}
\author{H. Shiromaru}
\affiliation{Department of Chemistry, Tokyo Metropolitan University, 1-1 Minamiosawa, Hachiouji-shi, Tokyo 192-0397, Japan}
\author{A. Cassimi}
\affiliation{CIMAP, CEA - CNRS - ENSICAEN, BP 5133, F-14070, Caen cedex 5, France}

\date{\today}

\begin{abstract}
Electron capture processes for low energy Ar$^{9+}$ ions colliding on Ar$_2$ dimer targets are investigated, focusing attention on charge sharing as a function of molecule orientation and impact parameter. A preference in charge-asymmetric dissociation channels is observed, with a  strong correlation between the projectile scattering angle and the molecular ion orientation. The measurements provide here clear evidences that projectiles distinguish each atom in the target and, that electron capture from near-site atom is favored. Monte Carlo calculations based on the classical over-the-barrier model, with dimer targets represented as two independent atoms, are compared to the data. They give a new insight into the dynamics of the collision by providing, for the different electron capture channels, the two-dimensional probability maps $p(\vec{b})$, where $\vec{b}$ is the impact parameter vector in the molecular frame.

\end{abstract}


\maketitle
Experiment on an elementary reaction with well-defined geometry, 
where the orientation of the reactant is fixed
in space and the impact parameter is well controlled is a challenging
subject which must considerably deepen the understanding
of chemical reactions. By virtue of molecule orientation 
techniques \cite{Kramer65,Sakai03}, or by event-by-event measurements
allowing to determine \textit{a posteriori}
the molecule orientation at the instant of the collision \cite{Horvat95,Werner97}, orientation dependence
in collisions involving highly charged ions has now been studied for a large 
energy range \cite{Horvat95,Werner97,Caraby97,Siegmann02,Ehrich02,Titze11,Kim14}.
Determination of the impact parameter is a more delicate issue: we cannot control the projectile beam and 
the target so as to collide with a desired impact parameter. 
Since Rutherford's gold foil experiment, access to the impact parameter is provided by the exchange of transverse momentum 
between the projectile and the target. But this transverse momentum exchange is usually very small,
and the emitted electrons contribute significantly to the momentum balance. Dependence on the molecule orientation may in that case
inform on the range of impact parameters involved in the collision process.
For fast collisions leading to multiple ionization of the target, orientation dependence 
was first qualitatively understood by a geometrical model \cite{Caraby97,Wohrer93}
and later interpreted using the Statistical Energy Deposition (SED) model \cite{Werner97,Siegmann02}.
In the specific case of dimer targets, molecular orientation dependence helped to the identification of one-step (one-site)
versus two-step (two-site) processes \cite{Titze11}, or allowed the determination of impact-parameter-dependent ionization
probability $p(b)$ for \textit{atomic} scattering processes \cite{Kim14}.

Now, when a diatomic molecule interacts with a projectile, one of the two atoms of the molecule
may lie closer to the projectile trajectory than the other. A charged projectile generates highly localized electric field, and
it is then natural to expect preferential ionization or capture from near-site atom to occur. 
To get access to such an atomic site sensitivity of the processes, one rely on the observation of both, the induced asymmetry 
in final charge sharing between the molecular fragments and, the orientation and position of the molecule with respect to the projectile ion trajectory. 
It requires a determination of the impact parameter \textit{vector} $\vec{b}$ in the molecular frame, only accessible through
measurements of transverse momentum exchange in coincidence with the molecule orientation.
For fast collisions, such a measurement seems out of reach.
But for multiple electron capture processes resulting from low energy highly-charged ion (HCI)-molecules collisions, the transverse momentum 
exchanged between the projectile and the molecule center-of-mass, without any emitted electron involved, may become measurable.

One difficulty remains in the fact that the
observed asymmetry is the final charge sharing on the molecular fragments and the latter
can be seriously disturbed by intramolecular charge redistribution.
For low-energy collisions with HCIs, there
are two conflicting scenarios for preferential charge sharing. 
To reduce the Coulomb repulsion
between the ionized target and projectile ion,
the electron sharing would lead to a lower charge on the
near atomic site. Such behavior has been observed for N$_2$ covalent molecules and very low energy collisions
(less than 100 eV/u), in an experiment sensitive to transverse momentum \cite{Ehrich02}.
Contrarily, if electron mobility in the
molecule is low, the near site would be preferentially ionized. 
In a previous paper, we have shown that the asymmetric charge sharing is favored
in the fragmentation of argon dimers multiply
ionized by electron capture in low energy collisions with Ar$^{9+}$ projectiles\cite{Matsumoto10}.
The preference of asymmetric sharing to symmetric one in
dimer case was interpreted in terms of ”low electron
mobility”. Provided that the charge sharing is practically
”frozen” in a dimer target, the measurement of the transverse momentum exchange combined with the measurement
of the dimer initial orientation should also show atomic site sensitivity in the outcome of the collision.
We present here an experimental and theoretical study of atomic site sensitivity for this collision system,
providing new insight into the multiple-electron capture processes for ion-molecule collisions.

The full details of the apparatus and data analyses are described in ref. \cite{Matsumoto10,Matsumoto11}.
Ar$^{9+}$ ions generated with an electron cyclotron resonance (ECR) ion source at the ARIBE-GANIL facility
(Caen, France) were accelerated to 15 qkeV and introduced to the collision
chamber. The Ar$_2$ dimer target was provided by a supersonic gas jet crossing the ion beam at 90$^\circ$.
Recoil ions resulting from charge transfer were collected using a uniform electric field
and detected by a microchannel plate with a delay-line
detector (DLD) giving both the time and
position of ion detection. Fragment ions from the dimers
were identified by double-hit time of flight (TOF) coincidence measurements
followed by a ”cleaning” procedure to eliminate false
coincidence events.
Detection of scattered projectile ions, which
generated a trigger for the TOF measurements of
recoil ions, was also position-sensitive, allowing to determine
the final charge state of the scattered projectile.
To determine the Kinetic Energy Release (KER) and orientation of the dissociating
dimer, the momentum of each fragment ion in the
center-of-mass coordinate was calculated from the position
and TOF data, imposing momentum conservation
restriction for optimal resolution \cite{Weber01}. 
The novelty, here, is that
the transverse momentum transferred to the dimer center-of-mass during
the collision was also inferred from these data sets, giving 
direct access to the transverse components of the scattered projectile momentum \cite{Cassimi96}.
In spite of a resolution limited by the finite size of the collision region, the scattering angle of the projectile
and its angle of emission, $\phi_{proj}$, 
was thus determined with reasonable accuracy.

The setup is only sensitive to charged fragments. The fragmentation channels 
that could be observed, 

Ar$_{2}^{2+}$ $\to$ Ar$^{+}$ + Ar$^{+}$ for double capture (DC),

Ar$_{2}^{3+}$ $\to$ Ar$^{2+}$ + Ar$^{+}$ for triple capture (TC),

Ar$_{2}^{4+}$ $\to$ Ar$^{3+}$ + Ar$^{+}$ 

and Ar$_{2}^{4+}$ $\to$ Ar$^{2+}$ + Ar$^{2+}$ for quadruple capture (QC),
are respectively denoted by (1,1)$_F$, (2,1)$_F$, (3,1)$_F$, and (2,2)$_F$.
To discriminate between fragmentation and capture channels, we use here the indices $F$ and $C$. For the double capture,
as previously shown in \cite{Matsumoto10,Matsumoto11},
the KER of the dissociating dimer gives access
to the electron capture multiplicity on each site of the dimer. We can thus distinguish between "two-site"
double capture (denoted here (1,1)$_C$) leading directly to coulomb explosion, and "one-site" double capture
(denoted here (2,0)$_C$) that can relax through radiative charge transfer (RCT).
For the fragmentation channels (2,1)$_F$ and (3,1)$_F$, the KER spectrum
shows no evidence of RCT  \cite{Matsumoto11}.
The transient non-dissociative molecular states populated by "one-site" TC and QC,
denoted respectively (3,0)$_C$ and (4,0)$_C$, lead respectively to the (2,1)$_F$ and (3,1)$_F$ fragmentation channels
through direct crossing with excited states. They can thus not be experimentally isolated from the (2,1)$_C$ and (3,1)$_C$ capture channels.
\begin{figure}
\includegraphics[width=\columnwidth]{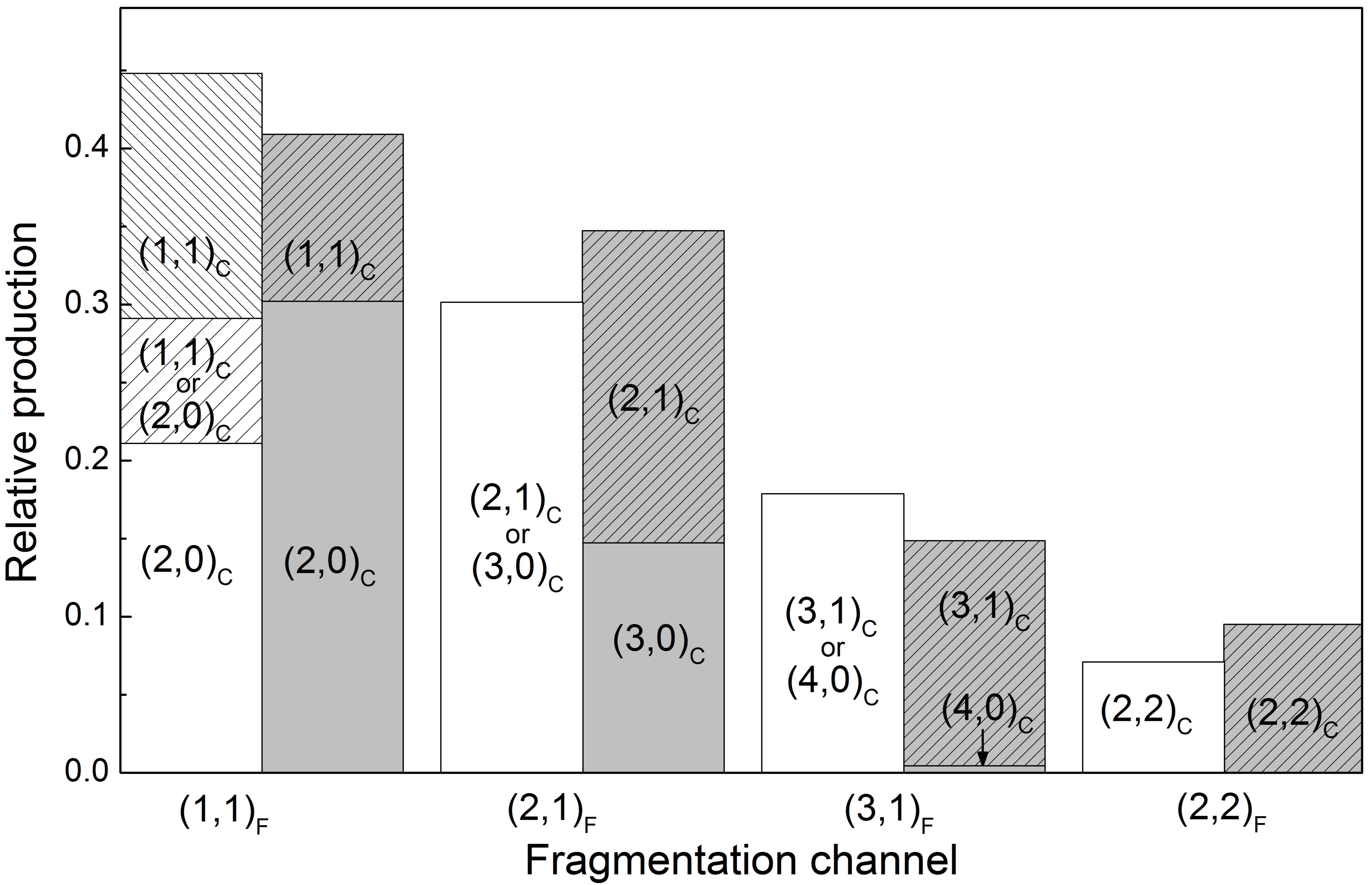}
\caption{ Relative yields for the different electron capture and fragmentation channels extracted from the experimental data (white) and results from MC COBM calculations (gray).}
\label{yields}
\end{figure}
The relative detected yields (without selection on the molecular orientation) associated to the fragmentation channels are shown in fig.1. 
The populations of the (2,0)$_C$ and (1,1)$_C$ channels
are extracted using a KER spectrum where both contributions overlap \cite{Matsumoto10}. About 20 \% of the
DC events  can thus not be clearly attributed to the (2,0)$_C$ or (1,1)$_C$ channels, as indicated in fig.1.
Relative uncertainties on all the fragmentation channel yields
are purely statistical and remain below 3 \%. The fig.1 clearly shows the preference for the asymmetric
fragmentation channels interpreted in terms of ”low electron mobility” between the two atoms of the dimer. 

In the present study, we focus attention onto
the angular correlation between the scattered projectile and the recoiling fragments for 
the different electron capture scenarios. 
To get a clearer view of impact parameter dependence in the molecular frame, we limit here the analysis to molecular targets oriented perpendicular 
to the beam axis at the moment of the collision. 
As schematically illustrated in the figures 2(a) and 2(b), the projectile scattering angle $\phi_{proj}$ is given by the direction
of the transverse momentum exchange due to the coulomb repulsion between the collision partners. 
It is thus closely related to the impact parameter vector $\vec{b}$ in the molecular frame and to the 
final charge on the two sites of the molecule.
In the fig.2(b), the molecular frame can be defined unambiguously
by the angle $\phi_{Ar^{3+}}$. In the following analysis, we will focus on the projectile scattering angle in the molecular frame,
$\phi_{diff}$=$\phi_{Ar^{A+}}-\phi_{proj}$, were $Ar^{A+}$ is the most charged fragment. 
For symmetric fragmentation channels, the two fragments of equal charge are not anymore distinguishable and
can both serve as a reference. This angle will in that case be defined $\textit{modulo}$ $\pi$.
\begin{figure}
\includegraphics[width=\columnwidth]{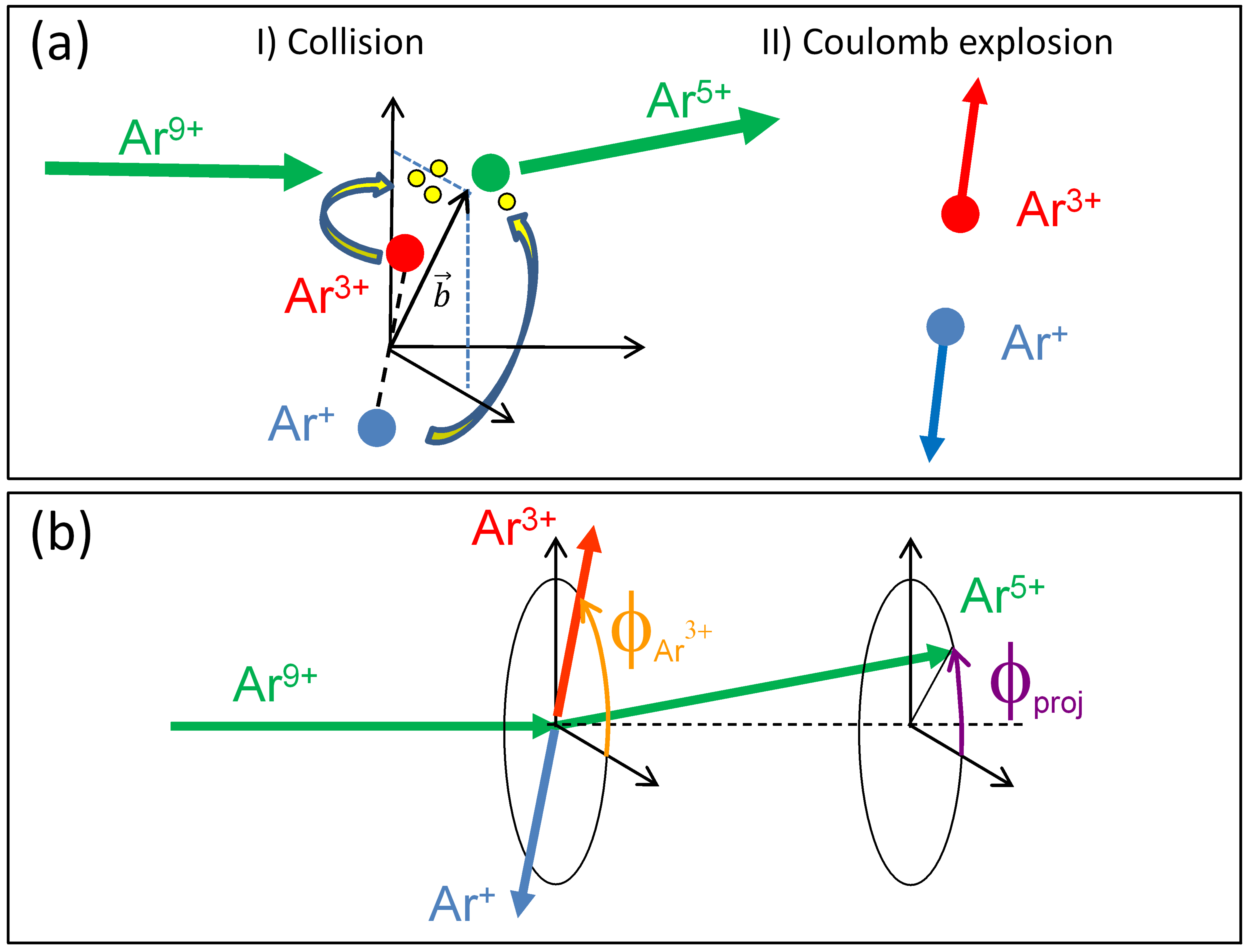}
\caption{(Color online) Schematic view of the multiple electron capture from Ar$_2$ by Ar$^{9+}$ projectiles
resulting in the (3,1)$_F$ asymmetric fragmentation channel (a). Representation of the scattering angle $\phi_{proj}$ and of the angle of emission
 $\phi_{Ar^{3+}}$ of the most charged fragment in the plane transverse to the beam axis (b).}
\label{orientation}
\end{figure}
The angular distributions in $\phi_{diff}$ of the capture channels of interest are shown in fig.3.
A selection of the data  corresponding to dimer targets oriented close to $90^\circ$ ($60^\circ-120^\circ$) in respect to the beam axis was first applied.
In spite of the low resolution obtained on the angle $\phi_{proj}$, one can see fig.3 that for the asymmetric channels (2,1)$_F$
and (3,1)$_F$ the projectile is preferentially scattered in the direction of the most charged fragment. In the simple picture of the collision
given by the fig.2, this is a clear evidence that electron capture from the near site is favored. 
The distributions of the (1,1)$_C$ and (2,2)$_F$ symmetric channels reach their maximum at $90^\circ$ and $270^\circ$, indicating  
dominant impact parameters close to the median plane of the dimer internuclear axis. For the channel (2,0)$_C$, 
we end up with a symmetric distribution: one cannot distinguish experimentally which fragment was initially ionized, prior the RCT decay process that
finally leads to the (1,1)$_F$ symmetric charge sharing. This loss of memory of the initial capture process leads to an angular distribution
quasi-isotropic. An asymmetry in the initial (2,0)$_C$ capture process can here only be clearly evidenced using calculations.
\begin{figure}
\includegraphics[width=\columnwidth]{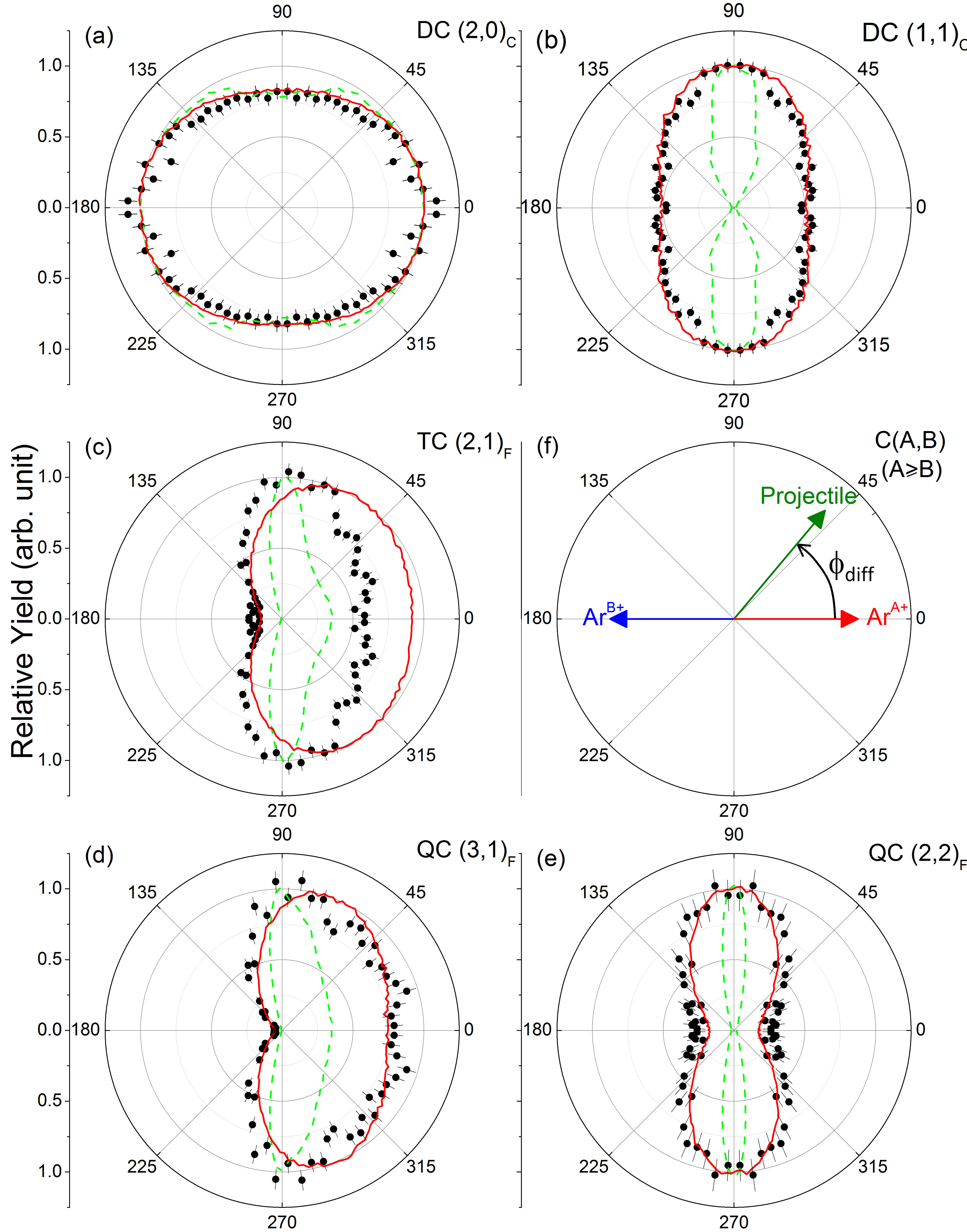}
\caption{(Color online) Angular distributions in $\phi_{diff}$ for the DC ((a) and (b)), TC (c), and QC ((d) and (e)) 
electron capture and associated fragmentation channels.
Experimental data (black dots) are compared to the calculations with (red lines), and without (dashed green lines) convolution with the 
experimental function response.}
\label{Phidiff}
\end{figure}

For low energy ion-atom collisions leading to multiple capture, the classical 
over-the-barrier model (COBM) \cite{Niehaus86} is known to be quite reliable.
An analytical theoretical treatment of the Ar$^{9+}+$$Ar_2$ collision based on the COBM and considering the dimer target as two Ar atoms fixed in space 
had already been performed \cite{Ohyama11}.  However,  the calculations
did not give access to projectile scattering angle and the "way out" of the collision was not yet included.  
We use here a different approach, also based on the COBM, but combined with Monte Carlo (MC) simulations.
The full MC COBM method and the present calculations will be further detailed in a forthcoming publication.
The orientation of the molecule with respect to the beam axis
as well as the impact parameter vector $\vec{b}$ in the molecular frame are sorted randomly. For each event,
the possible crossing points of the projectile trajectory with the capture radii of the two argon atoms are computed 
sequentially according to ref.\cite{Niehaus86} in the straight-line trajectory approximation.
If the two Ar atoms are here considered as fully independent, one site
can still indirectly influence the interaction between its neighbor and the projectile. In the "way out" of the collision,
this interaction can be affected by the possible change of the projectile charge that follows electron capture from the first site.
The sequential treatment allowed by a MC simulation is thus an essential feature of the model: along the projectile trajectory,
one can follow all the crossing points with capture radii and the subsequent charge sharing between the three partners of the collision.
The MC simulation approach also enables to compute the transverse momentum exchange
due to coulomb repulsion, and thus to determine the projectile scattering angle.
For this, we simply consider that the charge of the electrons shared by the projectile and the Ar atoms in the "way in" is distributed
between the different partners, while in the "way out" the charge of captured electrons is transferred from the target to the projectile.
Finally, the event-by-event mode of the MC COBM approach gives access to the correlation between initial conditions 
(impact parameter vector $\vec{b}$ and molecule orientation) and the outcome of the collision
(capture multiplicity on each site and projectile scattering angle). 
The relative yields obtained with this method for the different capture configurations are compared to the experimental results in fig.1.
As for the experimental data, calculations were first done accounting for all the molecular orientations. 
For "one-site" TC (3,0)$_C$ and QC (4,0)$_C$, 50\% of the populations given by the calculations 
can be statistically attributed to transient non-dissociative molecular states. They are thus respectively added to the final (2,1)$_F$ and (3,1)$_F$ fragmentation 
channels fed through direct crossings by these non-dissociative states.
Similarly, we also account for the 50\% of the DC (2,0)$_C$ population that dissociates prior RCT and does not end up in the (1,1)$_F$ fragmentation channel.
Without any adjustable parameter included in the model, calculations are found in very good agreement with the experimental data.

To be compared with the more detailed experimental results of fig.3, the angular distributions in $\phi_{diff}$ obtained 
with the MC COBM calculations have been convoluted with the instrumental resolution, limited by the of 0.6 mm (FWHM) diameter of the 
collision region. As for the experimental data, we have selected dimer targets orientations between $60^\circ$ and $120^\circ$.
For the channel (2,0)$_C$, the convoluted $\phi_{diff}$ distribution is displayed $\textit{modulo}$ $\pi$
to account for the RCT process and subsequent loss of information on the initial capture process.
After including these instrumental effects, calculations are once again found in excellent agreement with
experimental data for most of the capture or fragmentation channels. A substantial discrepancy can only be noticed for the (2,1)$_F$
channel, where the angular asymmetry predicted by the model is stronger than in the experimental data.
This could be simply explained here by an overestimation of the (3,0)$_C$ capture channel in the calculations.
\begin{figure}
\includegraphics[width=\columnwidth]{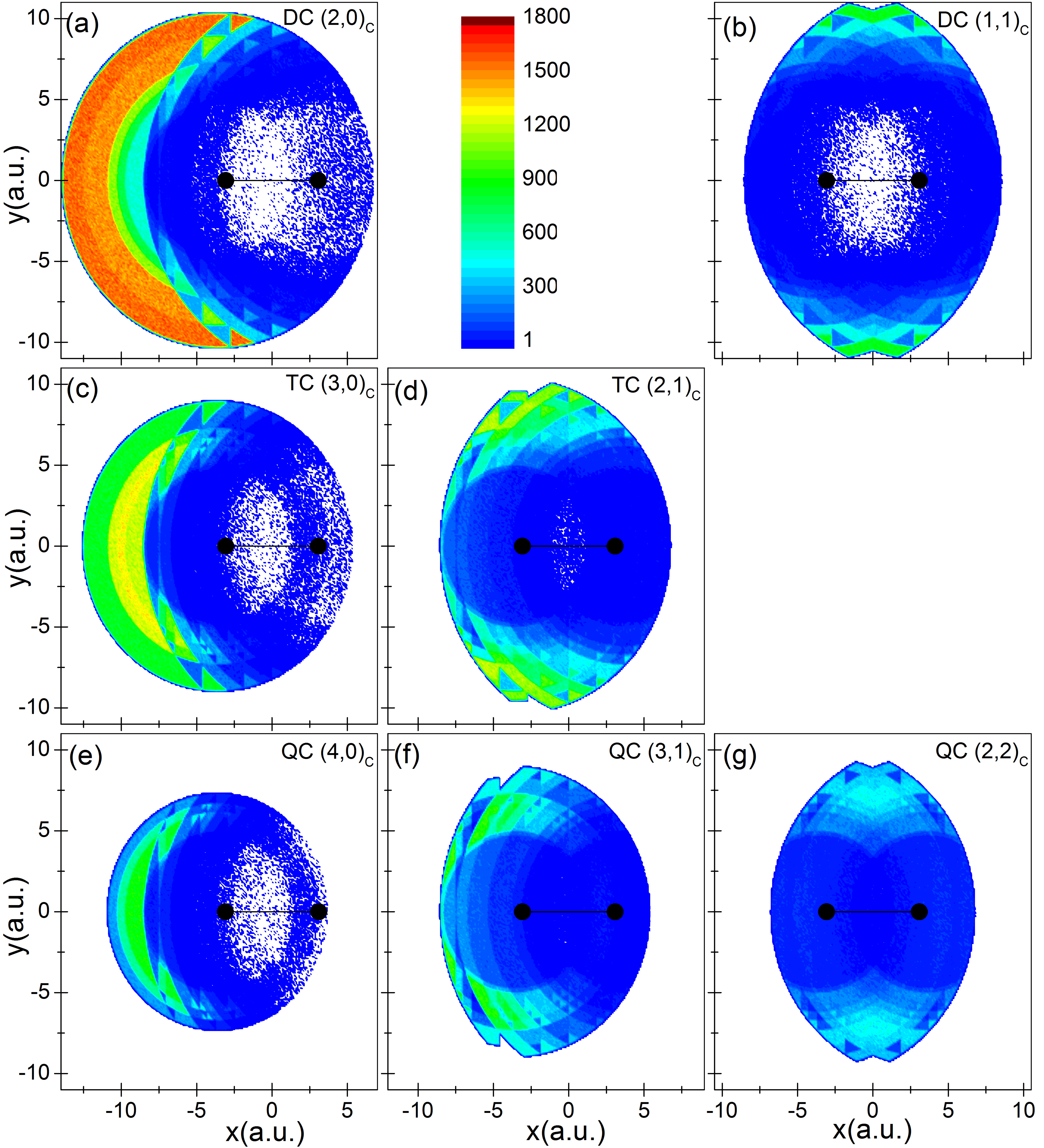}
\caption{(Color online) Representation in the molecular frame of the 2D maps $p(\vec{b})$ (in atomic units) associated with the different capture processes.
The positions of the two atoms are indicated by black dots. The atom with higher final charge state is on the left. The internuclear axis is chosen transverse to the 
projectile beam axis.}
\label{2Dmaps}
\end{figure}
To get a clearer view of the multiple capture process in the molecular frame, one can now directly look at the distributions
in impact parameter $p(\vec{b})$ leading to the different capture scenarios. The $p(\vec{b})$ distributions for dimer targets oriented at $90^\circ$ 
are displayed in the fig.4. For asymmetric capture configurations, and in particular for the (2,0)$_C$ one, we clearly confirm 
that capture from the near site is strongly favored. Impact parameters that contribute most to these processes cover a large area
on the side of the most charged fragment. For the symmetric capture channels, contributing impact parameters are restrained
close to the median plane of the internuclear axis. This partly explains the predominance of asymmetric fragmentation
channels on symmetric fragmentation channels. These calculations do not account for most of the complex
interactions and mechanisms at play in the multiple capture process and they cannot be perfectly accurate. Nevertheless,
the very good agreement between experimental data and calculations, in particular for the $\phi_{diff}$ distributions, attests that the model
gives here a realistic picture of the multiple capture process, even for structured targets such as rare gas dimers.

Measurements of the angular correlation between the scattered projectile and the recoiling 
fragments combined to model calculations have given access to atomic site sensitivity in low energy collisions between HCI and Ar$_2$ dimer.
They have shown that electron capture from "near-site" is strongly favored.
It is the opposite of what was previously observed with N$_2$ covalent molecules \cite{Ehrich02} and
may be a specific feature of rare gas dimer targets due to low electron mobility.
The same methodology could be now employed to investigate atomic site dependence for different projectile charges 
and for more complex targets, such as larger homonuclear or mixed clusters.

\bibliography{biblio}

\end{document}